\documentclass[a4paper,twoside]{article}

\usepackage{epsfig}
\usepackage{subcaption}
\usepackage{calc}
\usepackage{amssymb}
\usepackage{amstext}
\usepackage{amsmath}
\usepackage{amsthm}
\usepackage{multicol}
\usepackage{pslatex}
\usepackage{apalike}
\usepackage[bottom]{footmisc}
\usepackage{dsfont}
\usepackage[linesnumbered,ruled,vlined]{algorithm2e}
\usepackage{SCITEPRESS}     

\usepackage{xcolor}

\begin{document}

\title{Towards Programmable Memory Controller for Tensor Decomposition}

\author{
    Sasindu Wijeratne$^*$, Ta-Yang Wang$^*$, Rajgopal Kannan$^\dagger$, Viktor Prasanna$^*$
    \\ $^*$ University of Southern California, Los Angeles, USA
    \\ $^\dagger$US Army Research Lab, Los Angeles, USA \\ 
    Email: \{kangaram, tayangwa\}@usc.edu, rajgopal.kannan.civ@mail.mil, prasanna@usc.edu
    }


\abstract{Tensor decomposition has become an essential tool in many data science applications. Sparse Matricized Tensor Times Khatri-Rao Product (MTTKRP) is the pivotal kernel in tensor decomposition algorithms that decompose higher-order real-world large tensors into multiple matrices. Accelerating MTTKRP can speed up the tensor decomposition process immensely. Sparse MTTKRP is a challenging kernel to accelerate due to its irregular memory access characteristics. Implementing accelerators on Field Programmable Gate Array (FPGA) for kernels such as MTTKRP is attractive due to the energy efficiency and the inherent parallelism of FPGA. This paper explores the opportunities, key challenges, and an approach for designing a custom memory controller on FPGA for MTTKRP while exploring the parameter space of such a custom memory controller.}

\onecolumn \maketitle \normalsize

\section{\uppercase{Introduction}}
\label{sec:Introduction}

Recent advances in collecting and analyzing large datasets have led to the information being naturally represented as higher-order tensors. Tensor Decomposition transforms input tensors to a reduced latent space which can then be leveraged to learn salient features of the underlying data distribution. Tensor Decomposition has been successfully employed in many fields, including machine learning, signal processing, and network analysis~\cite{mondelli2019connection,cheng2020novel,wen2020tensor}. Canonical Polyadic Decomposition (CPD) is the most popular means of decomposing a tensor to a low-rank tensor decomposition model. It has become the standard tool for unsupervised multiway data analysis. The Matricized Tensor Times Khatri-Rao product (MTTKRP) kernel is known to be the computationally intensive kernel in CPD.
Since the real-world tensors are sparse, specialized hardware accelerators are becoming common means of improving compute efficiency of sparse tensor computations. But external memory access time has become the bottleneck due to irregular data access patterns in sparse MTTKRP operation.

Since real-world tensors are sparse, specialized hardware accelerators are attractive for improving the compute efficiency of sparse tensor computations. But external memory access time is the bottleneck due to irregular data access patterns in sparse MTTKRP operation.

Field Programmable Gate Arrays (FPGAs) are an attractive platform to accelerate CPD due to the vast inherent parallelism and energy efficiency FPGAs can offer. 
Since sparse MTTKRP is memory bound, improving the sustained memory bandwidth and latency between the compute units on the FPGA and the external DRAM memory can significantly reduce the MTTKRP compute time. FPGA facilitates near memory computing with custom adaptive hardware due to its reconfigurability and large on-chip BlockRAM memory \cite{xilinxalveo}. It enables the development of memory controllers and compute units specialized for specific data formats; such customization is not supported on CPU and GPU.

The key contributions of this paper are:
 \begin{itemize} 
\item We investigate possible sparse MTTKRP compute patterns and possible pitfalls while adapting sparse MTTKRP computation to FPGA.

\item We scrutinize the importance of a FPGA-based memory controller design to reduce the total memory access time of MTTKRP. Since MTTKRP on FPGA is a memory-bound operation, it leads to significant acceleration in total MTTKRP compute time.

\item We explore possible hardware solutions for Memory Controller design with memory modules (e.g., DMA controller and cache) that can use to reduce the overall memory access time. 
\end{itemize}

The rest of the paper is organized as follows: Section \ref{sec:Background} focuses on the background of tensor decomposition and spMTTKRP. Section \ref{sec:Tensor_Formats} and Section \ref{mttkrp_access_patterns} investigate the compute patterns and memory access patterns of spMTTKRP. Section \ref{sec:Memory_Controller_Requirements} discusses the properties of configurable Memory Controller design. Finally, we discuss the work in progress in Section \ref{discussion}.

\section{\uppercase{Background}}
\label{sec:Background}

\subsection{Tensor Decomposition}
\label{sec:TD}


Canonical Polyadic Decomposition (CPD) decomposes a tensor into a sum of 1D tensors~\cite{kolda2009tensor}. For example, it approximates a 3D tensor $\mathcal{X} \in \mathbb{R}^{I_0 \times I_1  \times I_2}$ as
\[
\mathcal{X} \approx \sum_{r=1}^R \lambda_r \cdot \mathbf{a}_r \otimes \mathbf{b}_r \otimes \mathbf{c}_r =: [\![ {\lambda} ; \mathbf{A}, \mathbf{B}, \mathbf{C} ]\!],
\]
where $R \in \mathbb{Z}_+$ is the rank, $\mathbf{a}_r \in \mathbb{R}^{I_0}$, $\mathbf{b}_r \in \mathbb{R}^{I_1}$, and $\mathbf{c}_r \in \mathbb{R}^{I_2}$ for $r=1, \ldots, R$. The components of the above summation can be expressed as factor matrices, i.e., $\mathbf{A} = [\mathbf{a}_1, \ldots, \mathbf{a}_R] \in \mathbb{R}^{I_0 \times R}$ and similar to $\mathbf{B}$ and $\mathbf{C}$. We normalize these vectors to unit length and store the norms in  $\lambda = [\lambda_1, \ldots, \lambda_R] \in \mathbb{R}^R$. Since the problem is non-convex and has no closed-form solution, existing methods for this optimization problem rely on iterative schemes. 

The Alternating Least Squares (ALS) algorithm is the most popular method for computing the CPD. Algorithm~\ref{cp-als} shows a common formulation of ALS for 3D tensors. In each iteration, each factor matrix is updated by fixing the other two; e.g., $\mathbf{A} \gets \mathcal{X}_{(0)}(\mathbf{B} \odot \mathbf{C})$. This Matricized Tensor-Times Khatri-Rao product (MTTKRP) is the most expensive kernel of ALS.

\begin{algorithm}
\DontPrintSemicolon
Input: A tensor $\mathcal{X} \in \mathbb{R}^{I_0 \times I_1 \times I_2}$, the rank $R \in \mathbb{Z}_+$ \;
Output: CP decomposition $[\![ {\lambda} ; \mathbf{A}, \mathbf{B}, \mathbf{C} ]\!]$, $\lambda \in \mathbb{R}^{R}$, $\mathbf{A} \in \mathbb{R}^{I_0 \times R}$, $\mathbf{B} \in \mathbb{R}^{I_1 \times R}$, $\mathbf{C} \in \mathbb{R}^{I_2 \times R}$ \;
\While{\emph{stopping criterion not met}}{
    $\mathbf{A} \gets \mathcal{X}_{(0)}(\mathbf{B} \odot \mathbf{C})$ \;
    $\mathbf{B} \gets \mathcal{X}_{(1)}(\mathbf{A} \odot \mathbf{C})$ \;
    $\mathbf{C} \gets \mathcal{X}_{(2)}(\mathbf{A} \odot \mathbf{B})$ \;
   Normalize $\mathbf{A}$, $\mathbf{B}$, $\mathbf{C}$ and store the norms as $\lambda$ \;
}
\caption{{\sc CP-ALS for the 3D tensors}}
\label{cp-als}
\end{algorithm}

\begin{figure}[h!]
\centering
\includegraphics[width=\linewidth]{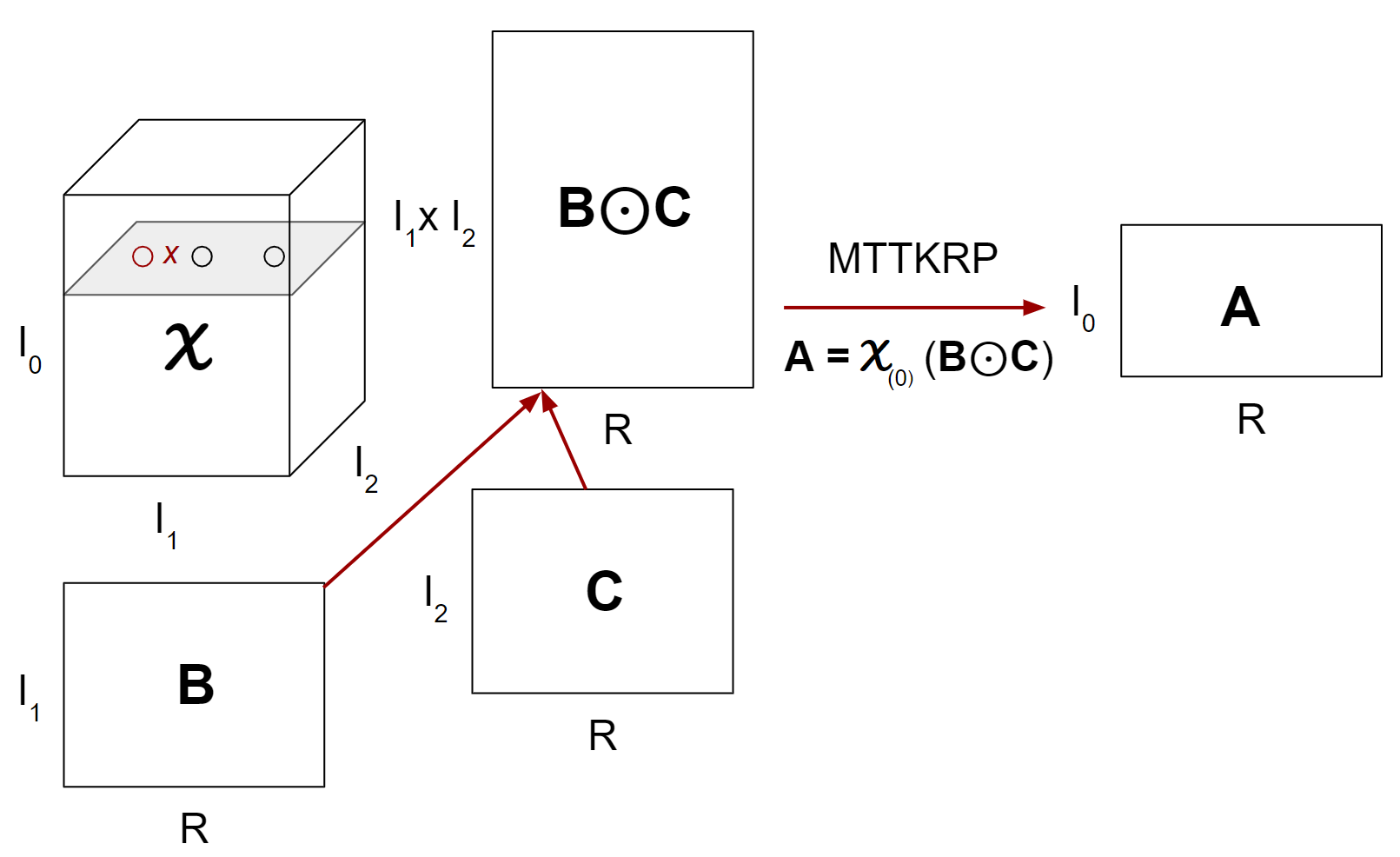}
\caption{An illustration of MTTKRP for a 3D tensor $\mathcal{X}$ in mode $0$}
\label{fig:mttkrp}
\end{figure}

Figure~\ref{fig:mttkrp} illustrates the update process of MTTKRP for mode $0$. With a sparse tensor stored in the coordinate format, sparse MTTKRP (spMTTKRP) for mode $0$ can be performed as shown in Algorithm~\ref{cp-als}. For each non-zero element $x$ in a sparse 3D tensor $\mathcal{X} \in \mathbb{R}^{I_0 \times I_1 \times I_2}$ at $(i, j, k)$, the $i$th row of $\mathbf{A}$ is updated as follows: the $j$th row of $\mathbf{B}$ and the $k$th row of $\mathbf{C}$ are fetched, and their Hadamard product is computed and scaled with the value of $x$. The main challenge for efficient computation is how to access the factor matrices and non-zero elements for the spMTTKRP operation. We will use a hypergraph model for modeling these dependencies in the spMTTKRP operation in Section~\ref{sec:Tensor_Formats}.

Algorithm \ref{gpu_paper_ref_mttkpr} shows the sequential sparse MTTKRP (spMTTKRP) approach for third-order tensors in Coordinate (COO) Format, where IndI, indJ, and indK correspond to the coordinate vectors of each non-zero tensor element. ``nnz" refers to the number of non-zero values inside the tensor.

\begin{algorithm}
\DontPrintSemicolon
\KwIn{indI[nnz], indJ[nnz], indK[nnz], vals[nnz], $\mathbf{B}[J][R]$, $\mathbf{C}[K][R]$}
\KwOut{$\mathbf{\tilde{A}}[I][R]$ }
\For{$z = 0 \emph{ to nnz}-1$}{
$i =$ indI[\textit{z}] \;
$j =$ indJ[\textit{z}] \;
$k =$ indK[\textit{z}] \;

\For{$r = 0 \emph{ to } R-1$}{
    $\mathbf{\tilde{A}}[i][r]$ += vals[\textit{z}] $\cdot$ $\mathbf{B}[j][r]$ $\cdot$ $\mathbf{C}[k][r]$ \;
}
}
return $\mathbf{\tilde{A}}$

\caption{{\sc Single iteration of COO based spMTTKRP for third order tensors}}
\label{gpu_paper_ref_mttkpr}
\end{algorithm}


\subsection{FPGA Technologies}
\label{sec:FPGA_Technologies}

FPGAs are especially suitable for accelerating memory-bound applications with irregular data accesses which require custom hardware. The logic cells on state-of-the-art FPGA devices consist of Look Up Tables (LUTs), multiplexers, and flip-flops. FPGA devices also have access to a large on-chip memory (BRAM). High-bandwidth interfaces to external memory can be implemented on FPGA. Current FPGAs are comprised of multiple Super Logic Regions (SLRs), where each SLR is connected to a single or several DRAMs using a memory interface IP.

HBM technology is used in state-of-the-art FPGAs as the high bandwidth interconnections particularly benefit FPGAs \cite{8916363}. The combination of high bandwidth access to large banks of memory from logic layers makes 3DIC architectures attractive for new approaches to computing, unconstrained by the memory wall. Cache Coherent Interconnect supports shared memory and cache coherency between the processor (CPU) and the accelerator. Both FPGA and the processor have access to shared memory in the form of external DRAM, while the cache coherency protocol ensures that any modifications to a local copy of the data in either device are visible to the other device. Protocols such as CXL \cite{CXL_web} and CCIX \cite{CCIX_web} develop to realize coherent memory.

\section{\uppercase{Sparse MTTKRP Compute patterns}}
\label{sec:Tensor_Formats}


The spMTTKRP operation for a given tensor can be represented using a hypergraph. For illustrative purposes, we consider a 3 mode sparse tensor $\mathcal{X} \in \mathbb{R}^{I_0 \times I_1  \times I_2}$ where $(i_0, i_1, i_2)$ denote the coordinates of $x$ in $\mathcal{X}$. Here $I_0$, $I_1$, and $I_2$ represent the size of each tensor mode. Note that the following approaches can be applied to tensors with any number of modes.

\begin{figure}[h!]
\centering
\includegraphics[width=\linewidth]{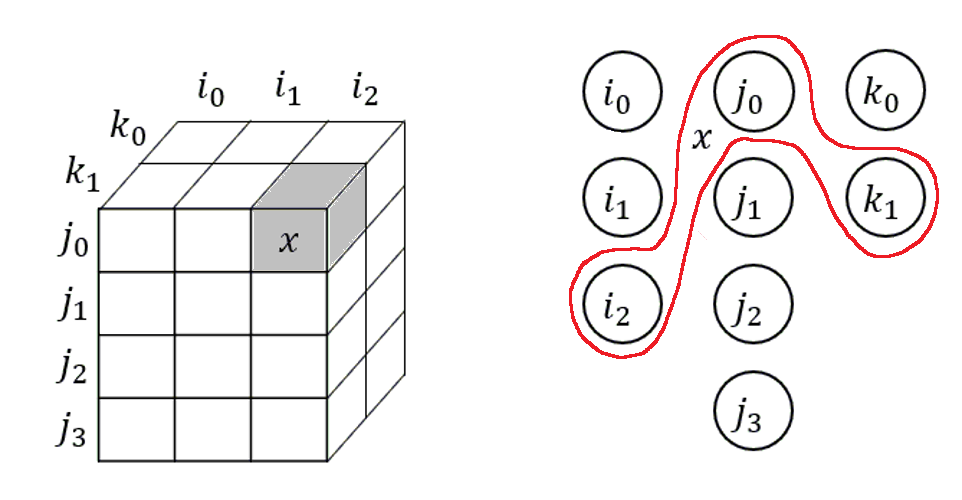}
\caption{A hypergraph example of a sparse tensor}
\label{hypergraph}
\end{figure}

For a given tensor $\mathcal{X}$, we can build a hypergraph $H = (V, E)$ with the vertex set $V$ and the hyperedge set $E$ as follows: vertices correspond to the tensor indices in all the modes and hyperedges represent its non-zero elements. For a 3D sparse tensor $\mathcal{X} \in \mathbb{R}^{I_0 \times I_1 \times I_2}$ with $M$ non-zero elements, its hypergraph $H = (V,E)$ consists of $|V| = I_0 + I_1 + I_2$ vertices and $|E| = M$ hyperedges. A hyperedge $\mathcal{X}(i, j, k)$ connects the three vertices $i$, $j$, and $k$, which correspond to the indices of rows of the factor matrices. Figure~\ref{hypergraph} shows an example of the hypergraph for a sparse tensor. 

Our goal is to determine a mapping of $\mathcal{X}$ into memory for each mode so that the total time spent on (1) loading tensor data from external memory, (2) loading input factor matrix data from the external memory, (3) storing output factor matrix data to the external memory, and (4) element-wise computation for each non-zero element of the tensor is minimized.



Considering the current works in the literature \cite{8735529} \cite{8821030} \cite{10.1109/SC.2018.00022} \cite{alto_paper}, for a given mode, there are 2 ways to perform sparse MTTKRP.
\begin{itemize}
\item Approach 1: Output-mode direction computation
\item Approach 2: Input-mode direction computation
\end{itemize}

Algorithm~\ref{mttkrp_appr_1_1} and Algorithm~\ref{mttkrp_appr_2} show the mode $0$ MTTKRP of a tensor with three modes using Approach 1 and Approach 2, respectively.

\begin{algorithm}
\DontPrintSemicolon
Input: A sparse tensor $\mathcal{X} \in \mathbb{R}^{I_0 \times I_1 \times I_2}$, dense factor matrices $\mathbf{{B}} \in \mathbb{R}^{I_1 \times R}$, $\mathbf{{C}} \in \mathbb{R}^{I_2 \times R}$ \;
Output: Updated dense factor matrix $\mathbf{A} \in \mathbb{R}^{I_0 \times R}$ \;

\For{each $i_0$ output factor matrix row in $\mathbf{A}$}{
$\mathbf{A}(i_0, :) = 0 $ \;
\For{each nonzero element in $\mathcal{X}$ at $(i_0,i_1,i_2)$ with $i_0$ coordinates}{
    Load($\mathcal{X}(i_0, i_1, i_2)$) \;
    Load($\mathbf{B}(i_1,:)$) \;
    Load($\mathbf{C}(i_2,:)$) \;
    \For{$r=1, \ldots, R$}{
        $\mathbf{A}(i_0, r) += \mathcal{X}(i_0, i_1, i_2) \times \mathbf{B}(i_1,r) \times \mathbf{C}(i_2,r)$ \;
    }
}
Store($\mathbf{A}(i_0,:)$) \;
}
\Return $\mathbf{{A}}$
\caption{{\sc Approach 1 for mode 0 of a tensor with 3 modes}}
\label{mttkrp_appr_1_1}
\end{algorithm}

\begin{algorithm}
\DontPrintSemicolon
Input: A sparse tensor $\mathcal{X} \in \mathbb{R}^{I_0 \times I_1 \times I_2}$, dense factor matrices $\mathbf{{B}} \in \mathbb{R}^{I_1 \times R}$, $\mathbf{{C}} \in \mathbb{R}^{I_2 \times R}$ \;
Output: Updated dense factor matrix $\mathbf{A} \in \mathbb{R}^{I_0 \times R}$ \;

\For{each $i_1$ input factor matrix row in $\mathbf{B}$}{
Load($\mathbf{B}(i_1,:)$) \;
\For{each nonzero element in $\mathcal{X}$ at $(i_0,i_1,i_2)$ with $i_i$ coordinates}{
    Load($\mathcal{X}(i_0, i_1, i_2)$) \;
    Load($\mathbf{C}(i_2,:)$) \;
    \For{$r=1, \ldots, R$}{
        $\mathbf{p_A}(i_0, r) = \mathcal{X}(i_0, i_1, i_2) \times \mathbf{B}(i_1,r) \times \mathbf{C}(i_2,r)$ \;
    }
    Store($\mathbf{p_A}(i_0, :)$) \;
}

\For{each $i_0$ output factor matrix row in $\mathbf{A}$}{
$\mathbf{A}(i_0,:) = 0$ \;
\For{each partial element $\mathbf{p_A}$ with $i_0$ coordinates}{
\For{$r=1, \ldots, R$}{
Load($\mathbf{p_A}(i_0, :)$) \;
    $\mathbf{A}(i_0, r) += \mathbf{p_A}(i_0, r)$
}
}
Store($\mathbf{A}(i_0, :)$) \;
}

}
\Return $\mathbf{{A}}$
\caption{{\sc Approach 2 for mode 0 of a tensor with 3 modes}}
\label{mttkrp_appr_2}
\end{algorithm}

We use the hypergraph model of the tensor to describe the different approaches. The main difference between these two approaches is the hypergraph traversal order. Hence, we denote the two approaches based on the order of hyperedge traversal. 

In Approach 1, all hyperedges that share the same vertex of the output mode are accessed consecutively. For each hyperedge, all the input vertices are traversed to access their corresponding rows of input factor matrices. In Approach 2, all hyperedges that share the same vertex of one of the input modes are accessed sequentially. For each vertex, all its incident hyperedges are iterated consecutively. For each hyperedge, the rest of the input vertices of the hyperedge are traversed to access rows of the remaining input factor matrices. It follows the element-wise multiplication and addition.

In Approach 1, since the order of hyperedge depends on the output mode, the output factor matrix can be calculated without generating intermediate partial sums (Algorithm~\ref{mttkrp_appr_1_1}: line 10). However, in Approach 2, since the hyperedges are ordered according to the input mode coordinates, we need to store the partial sums (Algorithm~\ref{mttkrp_appr_2}: line 9) in the FPGA external memory. It leads to accumulating the partial sums to generate the output factor matrix (Algorithm~\ref{mttkrp_appr_2}: line 11-17).


The total computations of both approaches are the same: for a general sparse tensor with $|T|$ non-zero elements, $N$ modes, and factor matrices with rank $R$, since every hyperedge will be traversed once, and there are $N-1$ multiplications and one addition for computing MTTKRP, the total computation per mode is $N \times |T| \times R$. However, their total external memory accesses are different: both approaches require $|T|$ load operations for all the hyperedges and the total factor matrix elements transferred per mode is $(N-1) \times |T| \times R$, which corresponds to accessing input factor matrices of vertices in the hypergraph model. However, in Approach 2, the partial value needs to be stored in the memory (Algorithm~\ref{mttkrp_appr_2}: line 10), which requires an additional $|T| \times R$ external memory storage. Let $I_{out}$ and $I_{in}$ represent the length of the output mode and the input mode, respectively. Then the total amount of data transferred is $|T| + (N-1) \times |T| \times R + I_{out}\times R$ for Approach 1 and $|T| + N \times |T| \times R + I_{in}\times R$ for Approach 2. Therefore, Approach 1 benefits from avoiding loading and storing partial sums. Table \ref{appr_properties} summarizes the properties of the two approaches.

\begin{table}[ht]
\caption{Comparison of the Approaches}
\begin{center}
\resizebox{\columnwidth}{!}{
\begingroup
\setlength{\tabcolsep}{6pt} 
\renewcommand{\arraystretch}{1.1} 
\begin{tabular}{ |c|c|c|c|c| }
 \hline
 \textbf{Approach} & \textbf{Total Computations} & \textbf{Total external memory accesses} & \textbf{Size of total partial sums} \\
 \hline\hline
1 & $N \times|T|\times R$ &  $|T| + (N-1) \times |T| \times R + I_{out}\times R$ & $0$ \\ 
 \hline
 2 & $N \times|T|\times R$ & $|T| + N \times |T| \times R + I_{in}\times R$ & $|T| \times R$ \\ 
 \hline
\end{tabular}
\endgroup
}
\label{appr_properties}
\end{center}
\end{table}


In the following, we discuss these in detail and identify the memory access characteristics. 

\subsection{spMTTKRP on FPGA} \label{FPGA_mttkrp_ii}

\begin{figure}
\centering
\includegraphics[width=\linewidth]{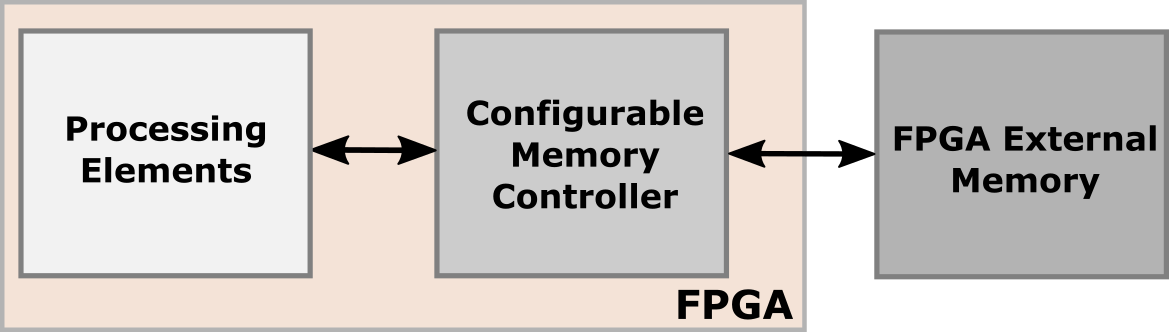}
\caption{Conceptual overall design}
\label{overall_arch}
\end{figure}

In this paper, we consider large-scale data decomposition on very large tensors.
Hence the FPGA stores the tensor and the factor matrices inside their external DRAM memory. Therefore, we need to optimize the FPGA memory controller to the DRAM technology. In this section, we first explain the DRAM timing model following the memory access patterns of sparse MTTKRP. Figure \ref{overall_arch} shows the conceptual overall design.

Approach 2 is not practical for FPGA due to the large external memory requirement to store the partial sums during the computation. In the work of this paper, we focus on Approach 1.

For Approach 1, the tensor is sorted according to the coordinates of the output mode. Typically, spMTTKRP is calculated for all the modes. To adapt Approach 1 to compute the factor matrices corresponding to all the modes, 
(1) Use multiple copies of the tensor. Each tensor copy is sorted according to the coordinates of a tensor mode or 
(2) re-order the tensor in the output direction before computing spMTTKRP for a mode.

Using multiple copies of a tensor is not a practical solution due to the limited external memory of the FPGA. Hence, our memory solution focuses on remapping the tensor in the output direction before computing spMTTKRP for a mode. It enables to perform spMTTKRP using Approach 1 for each tensor mode.

Algorithm \ref{mttkrp_appr_1} summarizes the Approach 1 with remapping. The algorithm focus on computing spMTTKRP of mode 1. Initially, we assume the sparse tensor is ordered according to the coordinates of mode 0 after computing the factor matrix of mode 0. Before starting the spMTTKRP for mode 1, we remap the according to the mode 1 coordinates (lines 3 - 6). After remapping, all the non-zero values with the same output mode coordinates are brought to the compute unit consecutively (line 9). For each non-zero value, the corresponding rows of the input factor matrices are brought into the compute units following the element-wise multiplication and addition. Since the tensor elements with the same output mode coordinates are brought together, the processing unit can calculate the output factor matrix without storing the partial values in FPGA external memory. Once a row of factor matrix is computed, the value is stored in the external memory. Here, Load/Store corresponds to loading/storing an element from the external memory. Also, ``$:$" refers to performing an operation for an entire factor matrix row.

The proposed approach introduces several implementation overheads:
\\
\textbf{Additional external memory accesses:}

The remapping required additional external memory load and a store (Algorithm \ref{mttkrp_appr_1}: lines 4 and 6). The total access to the external memory is increased by $2 \times |T|$ for a tensor of size $|T|$. The communication overhead per mode is: $$\dfrac{2 \times |T|}{|T| + (N-1) \times |T| \times R + I_{out}\times R}$$ $$\approx \dfrac{2}{1 + (N-1) \times R}$$

For a typical scenario (N = 3-5 and R = 16-64), the total external memory communication only increases by less than $6\%$.
\\
\textbf{Additional external memory space:}

During the remapping process, the remapped data requires an additional space equal to the size of the tensor ($|T|$) to store the remapped tensor elements in the memory. 

\textbf{Excessive memory address pointers to store the remapped tensor:} \label{track_me}

The remapping brings the tensor elements with the same output mode coordinate together (Algorithm \ref{mttkrp_appr_1}: line 5). To achieve this, the memory controller needs to track the memory location (i.e., memory address) of the next tensor element with each output coordinate needs to be stored. This required memory address pointers, which track the memory address to store a tensor element depending on its output mode coordinate. Algorithm \ref{mttkrp_appr_1} requires such memory pointers proportionate to the size of the output mode of a given tensor. 

The number of address pointers may not fit in the FPGA internal memory for a large tensor. For example, a tensor with an output mode with 10 million coordinate values requires 40 MB to store the memory address pointers (i.e., 32-bit memory addresses are considered). It does not fit in the FPGA on-chip memory. Hence the address pointers should be stored in the external memory. It introduces additional external memory access for each tensor element.

Also, the number of tensor elements with the same output mode coordinate value is different for each output coordinate due to the sparsity of the tensor. It complicates the memory layout of the tensor.

An ideal memory layout should guarantee:
(1) The number of memory address pointers required for remapping fit in the internal memory of the FPGA, and 
(2) Each tensor partition contains the same number of tensor elements.

\begin{algorithm}
\DontPrintSemicolon
Input: A sparse tensor $\mathcal{X} \in \mathbb{R}^{I_0 \times I_1 \times I_2}$ sorted in mode 0, dense factor matrices $\mathbf{{A}} \in \mathbb{R}^{I_0 \times R}$, $\mathbf{{C}} \in \mathbb{R}^{I_2 \times R}$ \;
Output: Updated dense factor matrix $\mathbf{B} \in \mathbb{R}^{I_1 \times R}$ \;

\For{each non-zero element in $\mathcal{X}$ at $(i_0,i_1,i_2)$ with $i_0$ coordinates}{
    Load($\mathcal{X}(i_0, i_1, i_2)$) \;
    pos$_{i_1}$ = Find(Memory address of $i_1$) \;
    Store($\mathcal{X}(i_0, i_1, i_2)$ at memory address pos$_{i_1}$) \;

}

\For{each $i_1$ output factor matrix row in $\mathbf{B}$}{
$\mathbf{B}(i_1, :) = 0 $ \;
\For{each non-zero element in $\mathcal{X}$ at $(i_0,i_1,i_2)$ with $i_1$ coordinates}{
    Load($\mathcal{X}(i_0, i_1, i_2)$) \;
    Load($\mathbf{A}(i_0,:)$) \;
    Load($\mathbf{C}(i_2,:)$) \;
    \For{$r=1, \ldots, R$}{
        $\mathbf{B}(i_1, r) += \mathcal{X}(i_0, i_1, i_2) \times \mathbf{A}(i_0,r) \times \mathbf{C}(i_2,r)$ \;
    }
}
Store($\mathbf{B}(i_1,:)$) \;
}
\Return $\mathbf{{B}}$
\caption{{\sc Approach 1 with remapping for mode 1 of a tensor with 3 modes}}
\label{mttkrp_appr_1}
\end{algorithm}


\section{\uppercase{Sparse MTTKRP memory access patterns}} \label{mttkrp_access_patterns}

The proposed sparse MTTKRP computation has 5 main actions: 
(1) load a non-zero tensor element, 
(2) load corresponding factor matrices, 
(3) perform spMTTKRP operation, 
(4) store remapped tensor elements, and 
(5) store the final output. 

The objective of the memory controller is to decrease the total DRAM memory access time. To identify the opportunities to reduce the memory access time, we analyze the memory access patterns of the proposed tensor format and memory layout. The summary of memory access patterns is as follows:

\begin{enumerate}
  \item The tensor elements can be loaded as streaming accesses while remapping and computing spMTTKRP.
\item Each remapped tensor element can be stored element-wise.
\item The different rows of each input factor matrices are random accesses.
\item Each row of output factor matrix can be stored in streaming memory access.
\end{enumerate}

Accessing the data in bulk (i.e., a large chunk of data stored sequentially) can reduce the total memory access time. It is due to the characteristics of the DRAM. DMA (Direct Memory Access) is the standard method to perform bulk memory transfers.
Further, the random accesses can be performed as element-wise memory accesses through a cache to explore the temporal and spatial locality of the accesses. It can improve the total access time.

Thus memory transfer types are as follows:

\begin{enumerate}
  \item \textbf{Cache transfers}: Supports random memory accesses. Load/store individual requests in minimum latency. Access patterns with high spatial and temporal locality are transferred using cache lines.
  \item \textbf{DMA stream transfers}: Supports streaming accesses. Load/store operations on all requested data with minimum latency from memory.
  \item \textbf{DMA element-wise transfers}: Transfer the requested data element-wise. This method is used with data with no spatial and temporal locality.
\end{enumerate}

\section{\uppercase{Towards Configurable Memory Controller}}
\label{sec:Memory_Controller_Requirements}

\begin{figure}
\centering
\includegraphics[width=\linewidth]{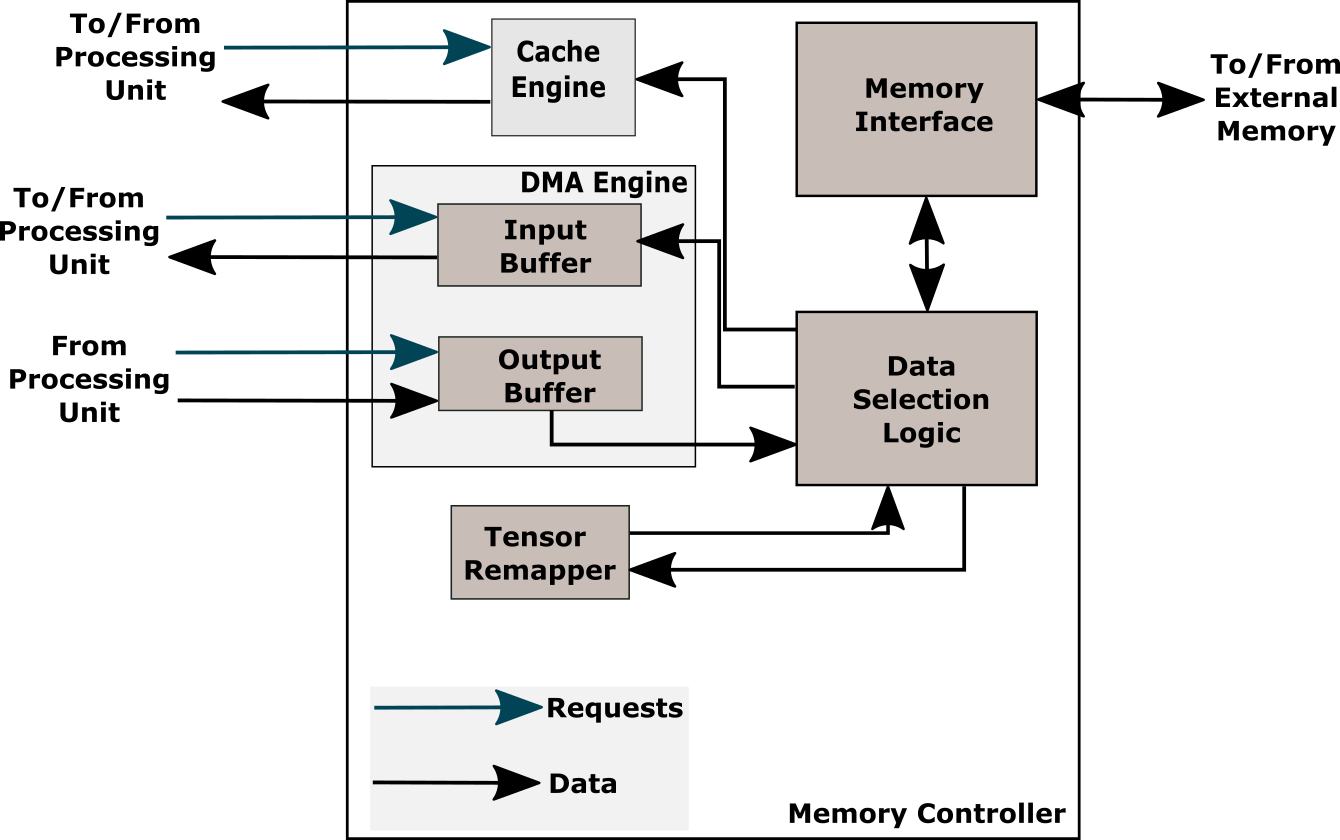}
\caption{Proposed Memory Controller}
\label{MC}
\end{figure}

To support the memory accesses, we propose a programmable memory controller as shown in Figure \ref{MC}. It consists of a Cache Engine, a tensor remapper, and a DMA Engine. We evaluate the impact of using caches and DMAs as intermediate buffering techniques to reduce the total execution time of sparse MTTKRP.

The modules inside the memory controller (e.g., Cache Engine, tensor remapper, and DMA Engine) can be developed as configurable hardware. These modules are programmable during the FPGA synthesis time. For example, the Cache Engine can have a different number of cache lines and associativity that can be configured during synthesis time. Also, the resource utilization of each module heavily depends on the configuration. On the other hand, the FPGA contains limited on-chip resources. Hence the FPGA resources should be distributed among different modules optimally, such that the overall memory access time is minimized (see Section \ref{DSE}).

\subsection{Memory Controller Architecture}

\subsubsection{Cache Engine} \label{sub_cache}
The Cache Engine can be used to satisfy a single memory request with minimum latency. The Cache Engine is used to explore the spatial and temporal locality of requested data. We intend to use Cache Engine to explore the locality of the input factor matrices. When a tensor computation requests a row of the factor matrix, the memory controller first look-ups the Cache Engine. If the requested factor matrix row is already in the Cache Engine due to prior requests, the factor matrix row is forwarded to the tensor computation from the Cache Engine. Otherwise, the tensor row is loaded to the Cache Engine from the FPGA external memory. Then a copy of the matrix row is forwarded to the computation while storing it in the cache.

\subsubsection{DMA Engine} \label{sub_dma}
The DMA Engine can process bulk transfers between the compute units inside FPGA and FPGA external memory. A DMA Engine can have several DMA buffers inside. 

The main advantages of having a DMA Engine are: 
(a) DMA requests can request more than one element at once, unlike the Cache Engine, and reduce the input traffic of the memory controller,
(b) Using a DMA Engine to access data without polluting the cache inside the Cache Engine, and
(c) DMA transfers can utilize the external memory bandwidth for bulk transfers.

\subsubsection{Tensor Remapper} 
Tensor remapper includes a DMA buffer and the proposed remapping logic discussed in Section \ref{sec:Tensor_Formats}. It loads each partition of the tensor as a bulk transfer similar to the DMA Engine. After, it stores each tensor element depending on the output mode coordinate value in an element-wise fashion.
\\\\
\textbf{Required memory consistency of the memory controller:}

The suggested memory controller above should have a weak memory consistency model with the following properties:

 \begin{itemize}

\item \textbf{Consistency of DMA Engine, Cache Engine and Tensor Remapper:} 
They process its requests based on a first-in-first-out basis.

\item \textbf{Consistency between Cache Engine, Tensor Remapper and DMA Engine: }
The first-in first-served basis is maintained. Since the same memory location is not accessed by the Cache Engine, Tensor Remapper and DMA Engine at the same time, weak consistency is maintained.
\end{itemize}

\subsection{Programmable Parameters}
The Cache Engine and DMA Engine use on-chip FPGA memory (i.e., BRAM and URAM). These resources need to be shared among the modules optimally to achieve significant improvements in memory access time. The resource requirement of the Cache Engine and DMA Engine depends on their configurable parameters mentioned below.

\subsubsection{Memory Controller Parameters} \label{cache_sub} \label{dma_sub}
Cache Engine parameters include cache line width, number of cache lines, and associativity of the cache.

The design parameters of the DMA Engine are: the number of DMAs, the number of DMA buffers per DMA, and the size of DMA buffers.

The design parameters of the Tensor Remapper include: 
(1) size of the DMA buffer, 
(2) width of a tensor element, and 
(3) the maximum number of address pointers Tensor Remapper can track.

\begin{table}
\caption{Characteristics of sparse tensors in FROSTT Repository}
\begin{center}
\resizebox{0.8\columnwidth}{!}{
\begingroup
\setlength{\tabcolsep}{6pt}
\renewcommand{\arraystretch}{1.2} 
\begin{tabular}{ |c|c|c|c| }
 \hline
\textbf{Metric} & \textbf{Value} \\ 
 \hline\hline
 Length of a tensor mode & $17$-$39$ M \\
 \hline
 Width of a matrix $(R)$ & $8 - 32$ (Typical = 16) \\
 \hline
 Number of non-zeros & $3$-$144$ M \\
 \hline
 Number of modes & $3$, $4$, $5$ \\
 \hline
 Tensor size & $\le 2.25$ GB \\
 \hline
 Size of a factor matrix & $< 4.9$ GB \\
 \hline
\end{tabular}
\endgroup
}
\label{summary}
\end{center}
\end{table}

\subsection{Exploring the Design Space} \label{DSE}

The tensor datasets can have different characteristics depending on the domain from which the dataset is extracted. Table \ref{summary} shows the characteristics of the tensors in The Formidable Repository of Open Sparse Tensors and Tools (FROSTT) \cite{frosttdataset}. It is commonly used in the high-performance computing community to benchmark custom accelerator designs for sparse MTTKRP.

Tensor datasets from separate domains of applications have different characteristics such as sparsity, size of the modes, and the number of modes. Hence, the datasets extracted from various applications show the least memory access time with different configurations of the memory controller.
Hence, performance estimator software is required to estimate the optimal configurable parameters for datasets of a domain. We introduce the features of a Performance Model Simulator (PMS) software to estimate the total execution time of spMTTKRP for a given dataset.
It can use with multiple datasets from the same domain to estimate the average execution time ($t_{avg}$) for a selected domain. Also, it should estimate the total FPGA on-chip memory requirement for a given set of programmable parameters to make sure the memory controller fits in the FPGA device. We will explore the possible inputs required for a PMS concerning:
(1) available FPGA resources (i.e., total BRAMs, and URAMs of the selected FPGA and data width of memory interface), 
(2) size of data structures (e.g., size of an input tensor element, size of an input factor matrix element, and rank of the input factor matrices), and
(3) Parameters of the memory controller (i.e., DMA buffer sizes, number of cache lines, associativity of cache, and number of factor matrices shared by a cache).

A module-by-module (e.g., Cache Engine and DMA Engine) exhaustive parameter search can be proposed to identify the optimal parameters for the memory controller.





\section{\uppercase{Discussion}}
\label{discussion}

In this paper, we investigated the characteristics of a custom memory controller that can reduce the total memory access time of sparse MTTKRP on FPGAs. 
Sparse MTTKRP is a memory-bound operation. It has 2 types of memory access patterns that can be optimized to reduce the total memory access time. A memory controller design that can be configured during compile/synthesis time depending on the application and targeted hardware is required.

We are developing a configurable memory controller and a memory layout for sparse tensors to reduce the total memory access time of sparse MTTKRP operation. 

Since synthesizing a FPGA can take a long time, optimizing the memory controller parameters for a given application can be a time-consuming process. Hence, we are developing a Performance Model Simulator (PMS) software to identify the optimal parameters for a given application on a selected FPGA.

\section*{ACKNOWLEDGEMENTS}

This work was supported by the U.S. National Science Foundation (NSF) under grants NSF SaTC \# 2104264 and PPoSS- 2119816.

\bibliographystyle{apalike}
{\small
\bibliography{example}}



\end{document}